\documentclass[a4paper,11pt]{article}
\usepackage{pos}
\usepackage{subfig}
\usepackage{floatrow}

\newfloatcommand{capbtabbox}{table}[][\FBwidth]

\title{Digital Quantum Simulation for Spectroscopy of Schwinger Model}
\ShortTitle{Digital Quantum Simulation for Spectroscopy of Schwinger Model}

\author*[a, b]{Dongwook Ghim}
\author[a, b]{Masazumi Honda}

\affiliation[a]{Interdisciplinary Theoretical and Mathematical Sciences Program (iTHEMS),
RIKEN, Wako, Saitama 351-0198, Japan}
\affiliation[b]{Yukawa Institute for Theoretical Physics (YITP), Kyoto University,
Sakyo-ku, Kyoto 606-8501, Japan}

\emailAdd{dongwook.ghim@riken.jp, masazumi.honda@riken.jp}

\abstract{This note discusses a method for computing the energy spectra of quantum field theory utilizing
digital quantum simulation. 
A quantum algorithm, called coherent imaging spectroscopy, quenches the vacuum with a time-oscillating perturbation and then reads off the excited energy levels from the loss in the vacuum-to-vacuum probability following the quench. 
As a practical demonstration, we apply
this algorithm to the (1+1)-dimensional quantum electrodynamics with a topological term known
as the Schwinger model, where the conventional Monte Carlo approach is practically inaccessible.
In particular, on a classical simulator, we prepare the vacuum of the Schwinger model on a lattice by adiabatic state preparation and then apply various types of quenches to the approximate vacuum
through Suzuki-Trotter time evolution. 
We discuss the dependence of the simulation results on the specific types of quenches and introduce various consistency checks, including the exact
diagonalization and the continuum limit extrapolation.
The estimation of the computational complexity required to obtain physically reasonable results implies that the method is likely efficient in the coming era of early fault-tolerant quantum computers.}

\FullConference{The 40th International Symposium on Lattice Field Theory (Lattice 2023)\\
July 31st - August 4th, 2023\\
Fermi National Accelerator Laboratory\\}

\begin{document}

\begin{flushright}
\texttt{ RIKEN-iTHEMS-Report-24 }
\end{flushright}

\renewcommand{\hookAfterAbstract}{%
    \par\bigskip
    }
\maketitle
	
\section{Introduction}
Recent technological advances on quantum computers have drawn 
the attention of the theoretical physics community, including the high-energy physics and lattice gauge theory community (e.g.~\cite{Bauer:2022hpo,Catterall:2022wjq,DiMeglio:2023nsa}).
The digital quantum simulation of quantum field theory is, in particular,
of interest because it naturally embeds the Hamiltonian formulation
of quantum field theory in its architecture \cite{Jordan:2012xnu, Jordan:2014tma}.
A great advantage of simulating quantum field theory in the Hamiltonian formulation over the conventional Monte Carlo approach is the absence of the infamous sign problem (e.g.~\cite{deForcrand:2009zkb,Aarts:2015tyj,Nagata:2021ugx}). 
Instead we typically have to deal with a huge vector space corresponding to the Hilbert space 
but one may be able to overcome that by utilizing quantum computer in future. 
Therefore it is worth extending and demonstrating the utility of quantum simulation in the context of high energy physics.

In this note, we will discuss the energy spectroscopy of field theory
as an application of quantum simulation to the problems in lattice gauge theories \cite{Banuls:2013jaa}.
Inspired by the experimental technique, called coherent imaging spectroscopy \cite{Senko:2014}, we provide a quantum algorithm that captures the 
energy eigenvalues of the lattice Hamiltonian. 
The key idea is to consider and simulate a dynamic process 
involving a state transition from the ground state to an excited state with control over the energy and frequency.

For the demonstration of our method,
we consider the Scwhinger model, $(1+1)$-dimensional quantum electrodynamics 
with non-trivial topological angle $\theta$ \cite{Schwinger:1962tp, Schwinger:1962tn}, which is a good test ground of quantum simulations in the context of high energy physics \cite{Martinez:2016yna,Muschik:2016tws,Klco:2018kyo,Kokail:2018eiw,Magnifico:2019kyj,Chakraborty:2020uhf,Yamamoto:2021vxp,Honda:2021aum,deJong:2021wsd,Honda:2021ovk,Tomiya:2022chr,Honda:2022hyu,Florio:2023dke,Lee:2023urk,Farrell:2023fgd,Nagano:2023uaq,Nagano:2023kge,Sakamoto:2023cxs,Angelides:2023noe}.
The Lagrangian density of the Schwinger model reads
\begin{align}
    \mathcal{L}_0 = \frac{1}{2g^2} F_{01}^2 + \frac{\theta}{2\pi} F_{01} + \overline{\psi} i \gamma^\mu \left( \partial_{\mu} + i A_{\mu} \right) \psi - m \overline{\psi} \psi \,,
\label{eq:con_naive}
\end{align}
where $m$, $g$, and $\theta$ stand for the mass of the electron, the coupling constant and
the topological angle, respectively. 
The two-component Dirac spinor associated with the 
electron is denoted by $\psi$ and the electric field strength is by $F_{01}$. 
Since the Schwinger model carries the non-trivial topological term 
in general, sign problem makes it hard to measure the observable with Monte Carlo sampling.
Thus, it is nice to perform its digital quantum simulation in order to unveil its physics which has not been captured with the aid of Monte Carlo techniques.

\section{Lattice formulation of the Schwinger model \label{formulation}}
To put the theory on quantum computer, we first put the Schwinger model on a lattice and map it to a spin system. 
Here, rather than directly working with 
\eqref{eq:con_naive},
we consider another equivalent Lagrangian obtained by the chiral rotation $\psi \rightarrow e^{ i\theta \gamma_5 /2}\psi$ to absorb the $\theta$-term as in \cite{Hamer:1997dx, Chakraborty:2020uhf}: 
\begin{align}
\mathcal{L}
= \frac{1}{2g^2} F_{01}^2 
+ \overline{\psi} i \gamma^\mu \left( \partial_{\mu} + i A_{\mu} \right) \psi
-m \overline{\psi} e^{i\theta\gamma^5}\psi ,
\end{align}
via the transformation of the path integral measure \cite{Fujikawa:1979ay}.
Then we put the theory on a lattice with 
open boundary condition.
In the temporal gauge, the lattice Hamiltonian in the staggered fermion formalism \cite{Kogut:1974ag, Susskind:1976jm} is given by
\begin{align} \label{H_chiral_bare}
    H= J \sum_{n=0}^{N-2} L_n^2 - i \sum_{n=0}^{N-2} \left( w - (-1)^n \frac{m_{\rm lat}}{2} \sin \theta \right) \left[ \chi_n^\dag U_n \chi_{n+1} - \textrm{h.c.} \right] + m_{\rm lat} \cos \theta \sum_{n=0}^{N-1} (-1)^n \chi_n^\dag \chi_n  \, ,
\end{align}
where $N$ is the number of lattice sites.
The lattice fields satisfy the commutation relations
\begin{align}
\{ \chi_n^\dag ,\chi_m \} = \delta_{mn},\ \ \{ \chi_n ,\chi_m \} =0 ,\ \  
[U_n ,L_m] =i \delta_{mn} U_n ,
\end{align}
and physical states are subject to the Gauss law: 
\begin{align}
L_n - L_{n-1} = \chi^\dag_n \chi_n - \frac{1- (-1)^n}{2}.
\end{align}
The parameters are defined in terms of lattice spacing $a$ and coupling constant $g$ as follows.
\begin{align}
J = \frac{g^2 a}{2} ,\quad  w = \frac{1}{2a} ,\quad m_\text{lat} = m - \frac{g^2}{16 w} ,
\end{align}
where we measure all the dimensionful quantities in the unit of $g$ and the last relation is according to \cite{Dempsey:2022nys, Dempsey:2023gib}.
Solving the Gauss law and applying the Jordan-Wigner transformation \cite{Jordan1928}:  
$\chi_n =  \Bigl(  \prod_{\ell <n }- i Z_\ell \Bigr) \frac{X_n -i Y_n}{2}$ 
with the Pauli spins $(X_n ,Y_n ,Z_n )$ at site $n$,
we obtain the following spin Hamiltonian
\begin{align}
    H = H_{ZZ} + H_{XX}+H_{YY} + H_{Z} ,
\end{align}
where 
\begin{align} \label{H_chiral}
\begin{split}
    H_{ZZ} & = \frac{J}{2} \sum_{n=1}^{N-2} \sum_{0 \leq k < \ell \leq n} Z_k Z_\ell = \frac{J}{2} \sum_{n=1}^{N-2} \sum_{k<n} (N-n-1) Z_k Z_n \,, \\
    H_{XX} &=\frac{1}{2}     \sum_{n=0}^{N-2} \left\{ w - (-1)^n \frac{m_{\rm lat}}{2} \sin \theta   \right\}  X_n X_{n+1}  \,, \\
    H_{YY} &= \frac{1}{2} \sum_{n=0}^{N-2} \left\{ w - (-1)^n \frac{m_{\rm lat}}{2} \sin \theta   \right\}  Y_n Y_{n+1}  \,, \\
    H_{Z} & = \frac{m_{\rm lat} \cos \theta}{2} \sum_{n=0}^{N-1} (-1)^{n} Z_n + \frac{J}{2} \sum_{n=0}^{N-2}  \textrm{mod}(n+1 , 2) 
    \sum_{\ell =0}^n Z_\ell \,.
\end{split}
\end{align}
This completes the qubit description of the Schwinger model.

\section{Simulation method for spectroscopy}
\begin{figure}[t]
\centering
\includegraphics[width=0.6 \textwidth]{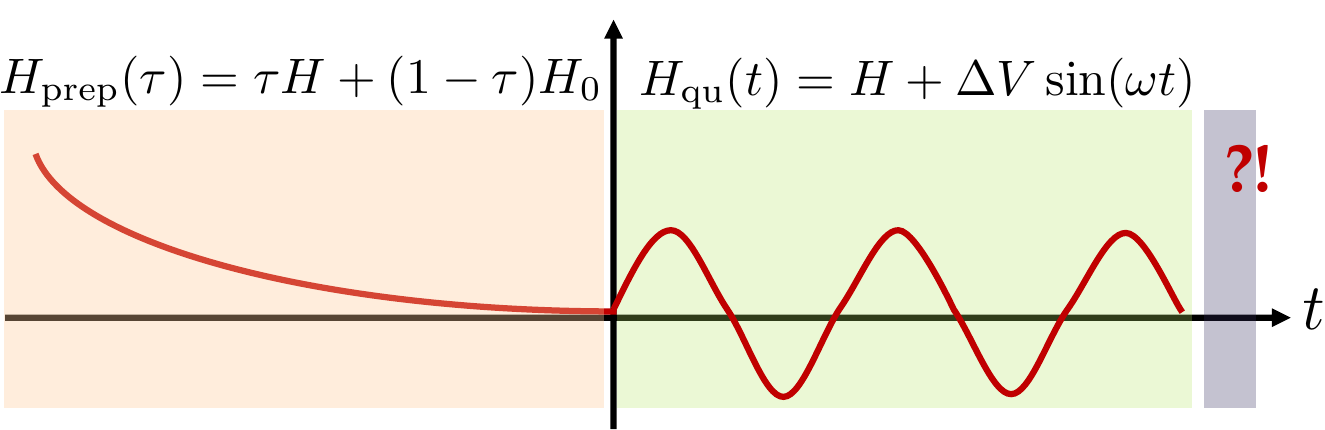}
 \caption{
 The cartoon of our simulation.
 The red line schematically represents the coefficients in the Hamiltonian.
 In the first stage of the simulation (orange),
 we ramp the coefficients so that the coefficients are dialed from the 
 initial Hamiltonian $H_0$
 to the target Hamiltonian \eqref{H_chiral}. 
 The next stage (green)
 simulates the sinusoidal oscillation of the parameters, either triggered by the pseudo-scalar condensate operator insertion or topological angle quench.
 At the end, 
 we measure the vacuum persistent probability.}
\label{fig:sketch}
\end{figure}

We first outline the simulation method for the spectroscopy of the lattice regularized theory. 
The main idea is to quench the ground state of the theory by an operator periodically oscillating in time with a particular frequency $\omega$ and measure the survival probability of the ground state.
If $\omega$ is close to the energy difference between one of excited states and the ground state, 
we have a transition to the excited state and the vacuum persistent probability becomes small at some time.
Repeating this for various values of $\omega$, one can estimate the energy spectrum.
In detail, we have freedom of choice in quantum algorithms at two moments: the ground state preparation and implementation of time evolution.  
In this note we simply adopt the adiabatic state preparation for the ground state and Suzuki-Trotter approximation for the time evolution \cite{Suzuki1991, Lloyd1073, Hatano:2005gh} while one could use different algorithms like variation-based ones depending on purposes.
The schematic cartoon of our simulation is drawn in Figure \ref{fig:sketch}.

Coming back to the Schwinger model, we prepare the ground state of the spin Hamiltonian \eqref{H_chiral} by adiabatically changing the initial state $ \left| \text{vac}_0 \right\rangle = \left| 1 0 1 0 \cdots 10 1 \right\rangle $,
which is the ground state of the simpler initial Hamiltonian: $H_0 := \left. H \right|_{w=0, \theta=0, m_{\rm lat}=M_0}$ with $M_0 >0$ as in \cite{Chakraborty:2020uhf, Honda:2021aum, Honda:2021ovk}.
By dialing the coefficients of terms in the Hamiltonian so that 
each coefficient interpolates the value in the initial Hamiltonian $H_0$
and the value in the target Hamiltonian \eqref{H_chiral},
we can prepare the ground state of the target Hamiltonian \eqref{H_chiral}.

Next, we introduce the gauge-invariant operator quench at a particular frequency to the prepared vacuum state.
Specifically, we will consider the pseudo-chiral condensate $ V = \int \overline{\psi} \gamma_5 \psi $.
With the spatial modulation introduced by $f_n$ taken into account, 
the quench on the lattice is translated into the following Pauli spin operators on qubits,
\begin{align} \label{H_condensate}
    \Delta H (t) = \frac{B_p}{2} \sum_{n=0}^{N-2} (-1)^{n+1} f_n \sin (\omega t) \left( X_n X_{n+1} + Y_n Y_{n+1} \right) \,.
\end{align}
The coefficient $B_p$ with the mass dimension 1 controls the strength of the external quench.

Besides the operator-type quench, time-sinusoidal fluctuation of physical parameters in the Hamiltonian
can also play the role of quench.
For instance, we consider the fluctuation in the topological angle, of which profile on the lattice is given in general,
\begin{align}
    \widetilde{\theta} (t,n) = \theta + \frac{B_p}{g} \, \delta \theta (t,n) = \theta + \frac{B_p}{g}  f_n \sin (\omega t) \,.
\label{th-fluct}
\end{align}
Such a parameter fluctuation is not a simple finite-term perturbation to the Hamiltonian. 
But its effect on the Hamiltonian can be
expanded in terms of the strength $B_p$, and the leading term reads
\begin{align} \label{H_theta}
\begin{split}
    \Delta H(t) &= \frac{m_\text{lat} B_p}{4g} \cos \theta \sum_{n=0}^{N-2} (-1)^n \, f_n \sin (\omega t) \left( X_n X_{n+1} + Y_n Y_{n+1} \right) \\
   &~~~ + \frac{m_\text{lat} B_p}{2g} \sin \theta \sum_{n=0}^{N-1} (-1)^n \, f_n \sin (\omega t) \, Z_n  +\mathcal{O}(B_p^2) \,.
\end{split}
\end{align}

The spatial modulation factor can decorate the quench by considering
site-dependent function $f_n$ in \eqref{H_condensate}.
A canonical choice on the basis function of spatial modulation is 
\begin{align}
    \left\{ f^{(k)}_n \right\}_{k=0,1, 2\cdots} \equiv \left\{ \cos \left( \frac{k \pi n}{N-1} \right) \right\} \,, 
\label{mod-basis}
\end{align}
which is the discrete version of $\{ \mathfrak{f}^{(k)} \, | \, \mathfrak{f}^{(k)} (x) =  \cos \left( \frac{\pi k x}{L} \right)  \text{ for } k= 0, 1, 2, \cdots \} $ .  
We 
call the integer $k$ above
as the mode number. 

We adopt the 2nd-order Suzuki-Trotter approximation in implementing time evolution onto the quantum circuit. In other words, we decompose the unitary time evolution operator triggered by Hamiltonian \eqref{H_chiral} at each Trotter step $\Delta t_{ST}$ as follows,
\begin{align}
    e^{-i \Delta t_{ST} H } \simeq e^{-i \frac{\Delta t_{ST}}{2} H_{XX} } e^{-i \frac{\Delta t_{ST}}{2} H_{YY} } e^{-i \Delta t_{ST} ( H_{ZZ} + H_Z ) } e^{-i \frac{\Delta t_{ST}}{2} H_{XX} } e^{-i \frac{\Delta t_{ST}}{2} H_{YY} } + \mathcal{O} ( 1/M^3 ) \,,
\end{align}
where an integer $ M $ stands for the total number of time steps during the quench simulation. Under the quench, we tune the coefficient of Hamiltonian \eqref{H_chiral} following either \eqref{H_condensate} or \eqref{th-fluct}.
Finally, we carry the measurement of the vacuum persistent probability
\footnote{In \cite{Senko:2014}, the authors used the various energy eigenstates $\langle E_n |$ 
for the basis of measurement following the quench. This was possible because they considered a Hamiltonian of which the form of eigenstate is well-known.
In a more generic situation, such information is not available and here we just observe the loss of ground state.
} 
\begin{align}
 \left| \langle \text{vac} | e^{- i \int dt ( H + \Delta H (t) ) } | \text{vac} \rangle \right|^2 \,,
\end{align}
after the quench. Technically, the last measurement procedure requires 
the adiabatic preparation of vacuum on the bra vector $ \langle \text{vac} | $
at the end of the quantum circuit for the simulation.

\subsection{Parameter set-up \label{sec:setup}}
\begin{figure}[t]
\centering
\includegraphics[width=0.5 \textwidth]{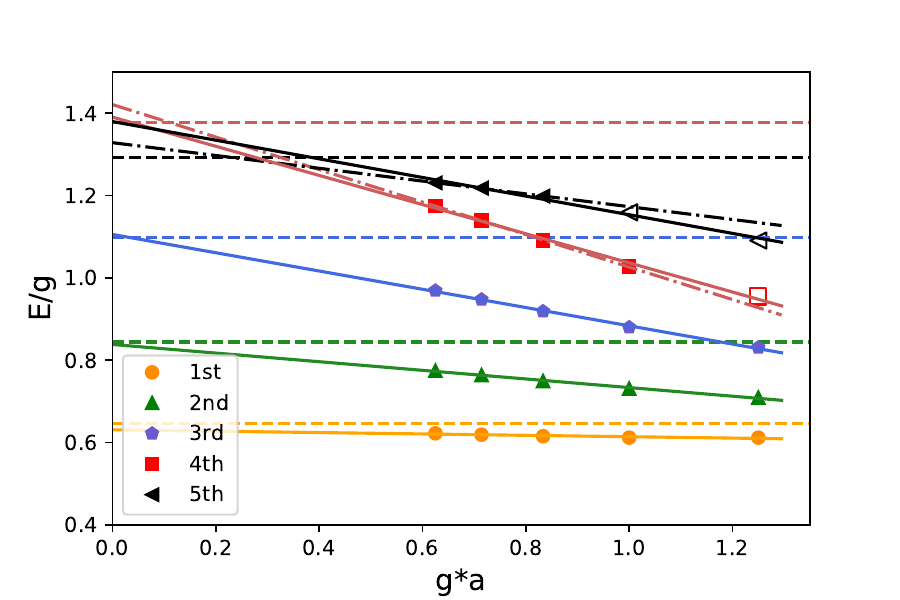}
 \caption{The five low-lying excited spectra of the Schwinger model with $gL=10 = (N-1)ga$ for $N=9 \,, 11 \,, 13 \,, 15 \,,$ and $17$ qubits, obtained by exact diagonalization. 
 The dashed horizontal lines are drawn at the analytic energy level of massless Schwinger model on the interval; $E_n^2 = M_S^2 + (n \pi /L)^2$ for $n=1, 2, 3, 4$
 where $M_S = g/\sqrt{\pi}$ and $E_{(2)} = 2 E_1$ for black dashed horizontal line. 
 The solid lines denote the linear fits over the five data points for each color while the two dashed-dotted lines denote the fits for the selected data points. 
\label{fig:continuum}}
\end{figure}

This section discusses how we set up the parameters of digital quantum simulation of spectroscopy. 
Above all, we set the length scale parameters like the length of the spatial interval $L$, by confirming the agreement of analytic continuum spectra and the lattice result obtained by the exact diagonalization at the massless case with $\theta=0$.
Figure \ref{fig:continuum} shows 
the comparison between analytic spectra and the result of the exact diagonalization of the spin Hamiltonian, computed by python-based package \texttt{QuSpin} \cite{Weinberg:2017igw, Weinberg:2019rfm}, and its extrapolation to the continuum limit $ a \rightarrow 0 $. 
Based on the $a \rightarrow 0 $ extrapolation, the five low-lying excited spectra, obtained by the exact diagonalization, 
are believed to capture the four different one-particle states of Schwinger meson (denoted in yellow, green, blue, and red) and
one two-particle excitation (marked in black)
at the interval length choice $gL=10$.

Before setting up the hierarchy among temporal scales of the quantum simulation, we identify four different kinematic (angular) frequency scales:
(i) Trotterization frequency $\omega_{ST} = \frac{2 \pi}{ \Delta t_{ST} }$ from Suzuki-Trotter approximation, (ii) quench frequency 
$ \omega \sim \Delta E_\text{gap}  $,  
(iii) resolution in frequency domain $\Delta \omega $,
and (iv) simulation time scale $T = M \Delta t_{ST} $ and its scale in frequency domain \textbf{$\Omega$} where $M$ stands for the number of Trotterization steps.
The perturbation theory with respect to $B_p$ suggests a useful dimensionless quantity $\gamma := \left| \langle f \left| \Delta V \right| \text{vac} \rangle \right| $ where $\Delta V$ is defined by the relation $\Delta H (t) = B_p \Delta V \sin (\omega t)$.%
\footnote{Recall that a tunable parameter $B_p$ controls the strength of quench and that it is a dimensionful parameter of which mass dimension is $1$, same to gauge coupling.} 
The brackets $ | \text{vac} \rangle $ and $ \langle f | $ stand for the initial vacuum state and target excited energy eigenstate, respectively.

The perturbation theory \cite{Dirac:1927dy} says that the transition probability between the vacuum and target state $| f \rangle $ whose energy gap is $\Delta E_\textrm{gap}$ is given by
\begin{align}
    P_{\text{vac} \rightarrow f} (t)
    = \left( \gamma B_p \right) ^2 \frac{ \sin^2 \left[ \left( \Delta E_\text{gap} - \omega \right) t \right] }{\left( \Delta E_\text{gap} - \omega \right)^2 } 
    + \mathcal{O} \left( B_p^3 \right ) \,.
\label{transition-probability}
\end{align} 
To proceed with the further analysis based on the perturbation theory, let us introduce two assumptions on the dynamical process of state transition and its simulation: (i) a state transition occurs in a short enough simulation time in which the perturbation theory is valid,
(ii) a window of quench frequency $\Delta \omega$ is fine enough so that the argument inside the $\sin$ function in the denominator of \eqref{transition-probability} is small enough, \textit{i.e.} $\Delta \omega \, T < 1 \,,$ near the energy gap $ \omega \sim \Delta E_\textrm{gap} $.

Now, let us set the probability threshold $P_\textrm{th}$ such that $0 \lesssim P_\textrm{th} < 1$ as follows. 
When the vacuum persistent probability after the quench is smaller than $1-P_\textrm{th} $, 
we identify the loss of vacuum so read the frequency of quench as the energy gap.
Then, the lower bound of estimated simulation time $T_\textrm{th} $ reads in terms of characteristic scale and preset parameters;
$ T > T_{\textrm{th}} = \frac{ \sqrt{P_\textrm{th} } }{ \gamma B_p } \,. $

\begin{table}[t]
\begin{center}
\scriptsize
\begin{tabular}{|c  c   c c  | c c c c |}  
 \hline
\textbf{parameters} & \textbf{symbol} & \textbf{value} & \textbf{remark}  & 
\textbf{parameters} & \textbf{symbol} & \textbf{value} & \textbf{remark} \\ 
 \hline
 \hline
 the number of qubits & $N$ & 9 & odd integer &  the length of interval & $L$ & $ 10.000 $  & $ L=(N-1)a $  \\
 lattice spacing & $a$ & 1.250 & & the inverse lattice constant & $w$ & $ 0.400 $ & $w=1/(2a)$ \\
 the number of shots & $N_s$ & $2000$ & & quench frequency gap & $\Delta \omega$ & $ 0.050 $ &   \\
 the number of steps & $M$ & 2000 & 
 & steps for adiabatic preparation & $M_\text{adia}$ & 2500 & $M_{\text{adia}}/M = 1.25 $ \\
 simulation time & $T$ & $ 73.000 $ &  & IR frequency cutoff & $\Omega$ & $ 0.086 $ & $\Omega =  2 \pi /T $ \\
 Trotterization time & $\Delta t_{ST} $ & $ 0.0365  $ & $\Delta t_{ST} =T/M$ & 
 Trotteriztation frequency & $\omega_{ST}$ & $ 1.720 $ & $ \omega_{ST} = 2 \pi / \Delta t_{ST} $  \\
 \hline
\end{tabular}
\end{center}
\caption{The choice of the simulation parameters. 
\label{tab:parameters}}
\end{table}

The resolution of probe frequency $\Delta \omega$ should be smaller than the differences of the excitation energies, which are mostly attributed to the higher momentum modes of the particle states in quantum field theories.  
In the regime of small electron mass and small topological angle $m \simeq 0 \,, \theta \simeq 0 $, we have $\Delta \omega / \omega <  (\pi/L) / M_S $ where $M_S$ stands for the mass of dual scalar Schwinger meson at $\theta = 0$. The right-hand side of the inequality is $\mathcal{O}(1)$ in our simulation whose parameters are given in Table \ref{tab:parameters}.

For the validity of Suzuki-Trotter approximation, the following inequality should hold; $ \omega \ll \omega_{ST} \,.$
More accurately, the order of magnitude of accumulative error arising from 2nd order Suzuki-Trotter approximation scales as $ \epsilon_{ST} \sim \mathcal{O} \left( M \left( \omega \Delta t_{ST} \right)^3 \right) \sim \mathcal{O} \left( \omega^3  \omega_{ST}^{-2} \Omega^{-1} \right) \,.$
Thus, the reliable simulation requires the accumulative error $\epsilon_{ST}$ to be small
and we find
$ \omega_{ST} \gg \Omega^{- \frac{1}{2}} \omega^{\frac{3}{2}} \,.$
Synthesizing the scaling laws obtained above, we obtain the hierarchy between frequency scales; $ \Delta \omega < \Omega < \omega < \omega_{ST}  \,.$
This is consistent with the numerics given in Table \ref{tab:parameters}.

\section{The result of simulation}

This section presents the simulation results of spectroscopy using a classical simulator (IBM Qiskit) for two types of quenches: pseudo-chiral condensate quench 
and topological angle quench. 
Note that the momentum is not a conserved quantity as we impose the open boundary condition on both ends of the spatial interval.
On the other hand, the mode number, introduced with the spatial modulation in the form of \eqref{mod-basis},
turns out to label and distinguish the low-energy eigenstates.

\subsection{Pseudo-chiral condensate quench}

\begin{figure}[t]
\scriptsize
\subfloat[$k=0 $]{{\includegraphics[width=0.27 \textwidth ]{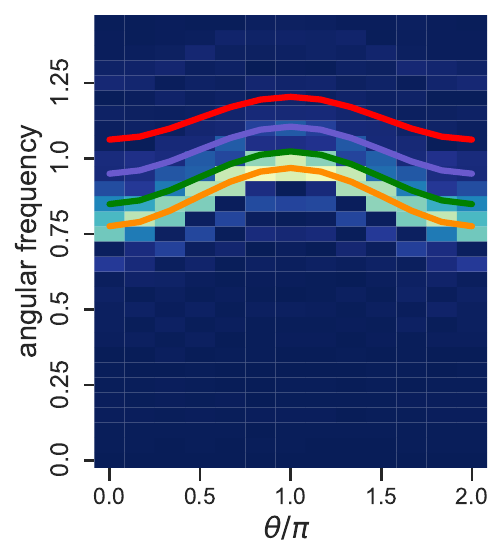} }}%
\subfloat[$ k=1 $]{{\includegraphics[ width=0.25 \textwidth]{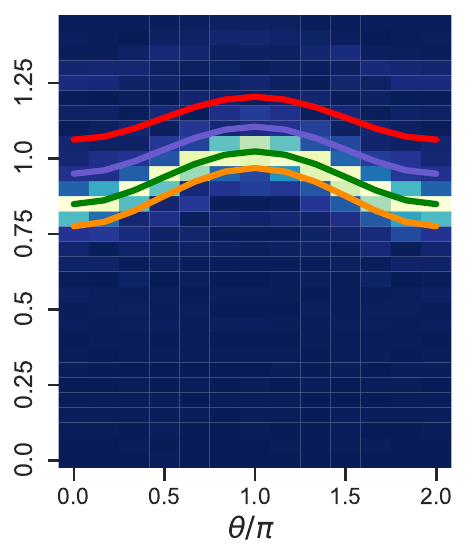} }}%
\subfloat[$ k=2 $]{{\includegraphics[width=0.25 \textwidth]{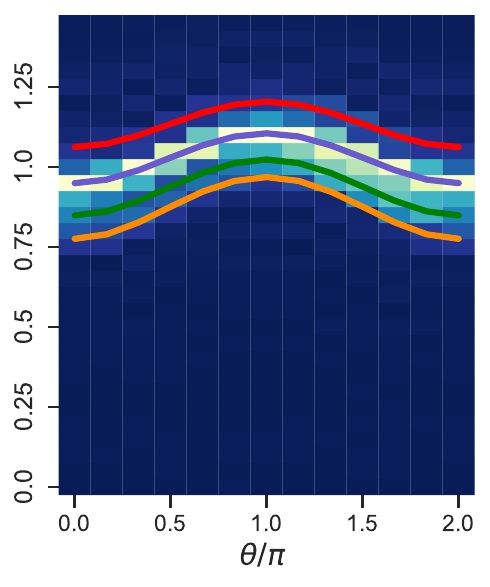} }}%
\subfloat[$ k=3 $]{{\includegraphics[width=0.31 \textwidth]{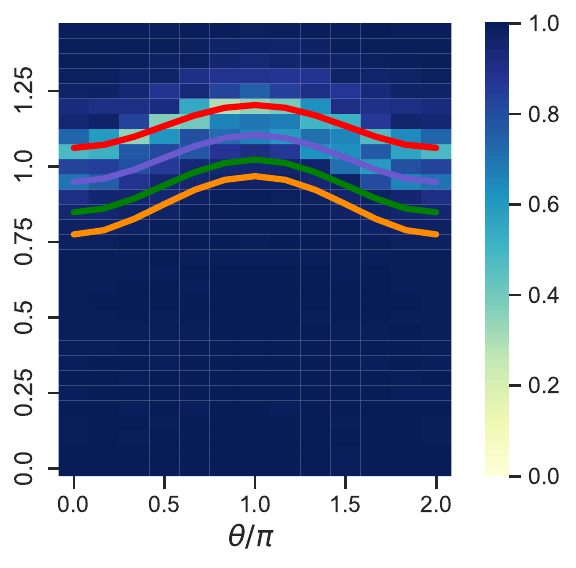} }}%
\caption{The density plot for the vacuum persistent probability 
for $m=0.100 $ and for the various topological angle under the pseudo-chiral
condensate quench $\Delta V = \int dx \, \mathfrak{f} (x) \, \overline{\psi} (x) \gamma_5 \psi (x) $. Solid lines denote the exact diagonalization result with \texttt{QuSpin}. 
The coefficient in the quench of the form \eqref{H_condensate} is chosen as $B_p = 0.011 \,.$
In each plot, the quench operator carries a distinct
mode number (a) $k=0$ , (b) $k = 1$ , (c) $ k=2 $, and (d) $ k=3 $ when the continuum limit of the modulation is defined by $\mathfrak{f} (x) = \cos (  k \pi x / L ) \,. $ }%
\label{fig:psi_quench_prob} %
\end{figure}

Figure \ref{fig:psi_quench_prob} shows the vacuum persistent probability under the pseudo-chiral condensate quench of particular frequency and spatial modulation 
at the 
electron mass $m = 0.1 $ and 
various values of topological angle $ \theta \in \{ 0 \,, \pi /6 \,, 2 \pi / 6 \,, \cdots \,, 2 \pi \} \,.$ 
Quench with the mode number $k$ induces the excitation to $(k+1)$-th lowest excited state.

When the topological angle vanishes, the intuition based on dual scalar Schwinger meson \cite{Schwinger:1962tn, Schwinger:1962tp}
tells that a low-energy excited
states are the higher-mode excitations of a single scalar particle, thus their energy is given
by $E_n^2 = M_S^2 + (n \pi /L)^2 $,
where $M_S = g/\sqrt{\pi}$.
Moreover, since each $n$-th excited state is excited by
the quench of mode number $(n-1)$,
one can interpret this in terms of
particle-in-a-box solution of Schr\"odinger equation. In such a quantum mechanics problem
of single particle, the wave function of
$n$-th excited state carries $(n-1)$ nodes,
the number of which is the same to the mode number necessary for the excitation of the corresponding state.
Through the simulation of small electron mass regime, 
the intuition at the vanishing topological angle
proves its applicability to the regime of the non-trivial topological angle,
since the quench of the same mode number successfully induces the excitation of the vacuum.

\subsection{Topological angle quench}

\begin{figure}[t]
\scriptsize
\subfloat[$ k =0 $]{{\includegraphics[width=0.27 \textwidth ]{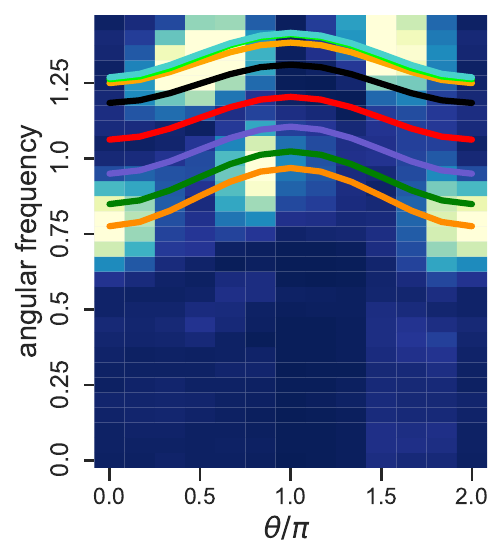} }}%
\subfloat[$ k= 1 $]{{\includegraphics[width=0.25 \textwidth ]{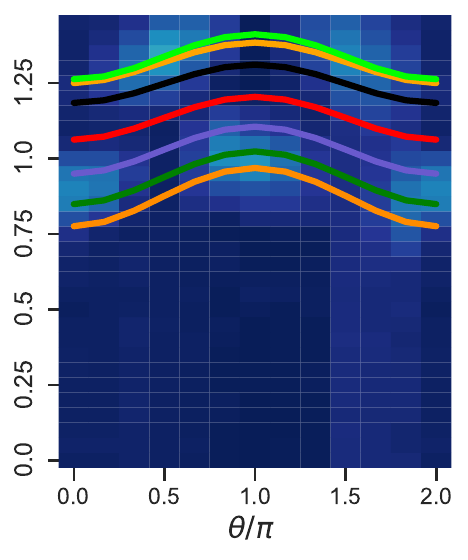} }}%
\subfloat[$ k=2 $]{{\includegraphics[width=0.25 \textwidth ]{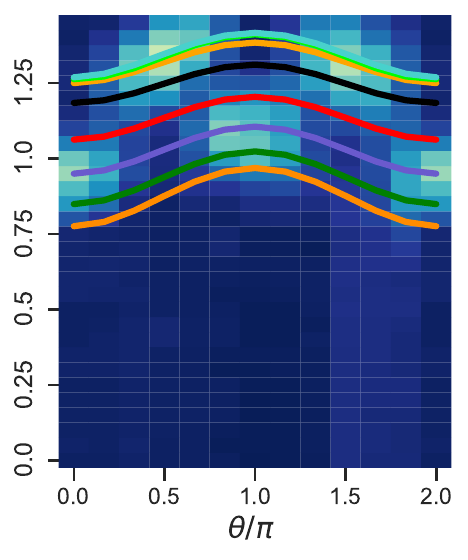} }}%
\subfloat[$ k=3 $]{{\includegraphics[width=0.31 \textwidth ]{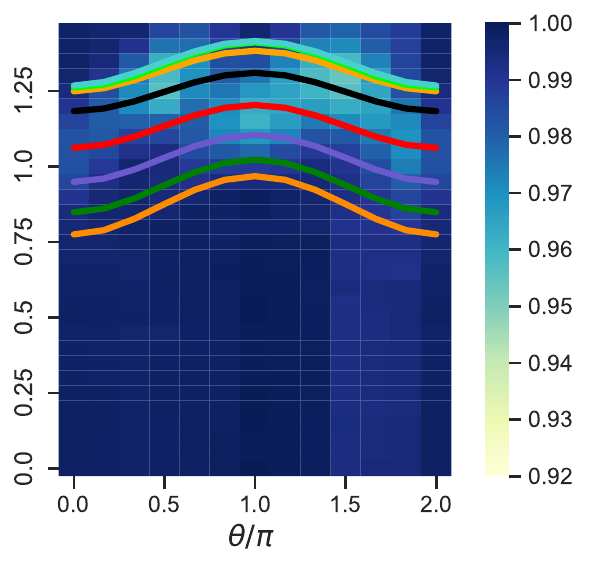} }}%
\caption{
A similar plot to Figure.~\ref{fig:psi_quench_prob} for the topological angle quench $\widetilde{\theta} (x) = \theta + (B_p /g ) \, \delta \theta (t, x)  $.
The coefficient in the quench of the form \eqref{th-fluct} is chosen as $B_p = 0.500 $.
}
\label{fig:th_quench_prob}%
\end{figure}

Figure \ref{fig:th_quench_prob} shows the density plot of the vacuum persistent probability after Trotterized time evolution with theta fluctuation quench \eqref{th-fluct}.
Unlike the pseudo-chiral condensate case, Figure \ref{fig:th_quench_prob} exhibits the excitations at higher energy near $\theta = \frac{\pi}{2}$ and $\frac{3 \pi}{2}$.
We suspect that they correspond to $2$-particle states under $\theta \rightarrow 0 $ limit
and the transition amplitude between the 2-particle state and the vacuum under the theta fluctuation is non-trivial whereas its counterpart amplitude with the pseudo-chiral condensate operator almost vanishes:
\begin{align}
\label{eq:selection}
    \left\langle \text{2-ptl st} \right|  
    \sum_{n=0}^{N-1} (-1)^n \,  Z_n 
    \left| \text{vac} \right\rangle \neq 0 \,, \quad
    \left\langle \text{2-ptl st} \right|  
    \sum_{n=0}^{N-1} (-1)^n & \, ( X_n X_{n+1} + Y_n Y_{n+1} )
    \left| \text{vac} \right\rangle \simeq 0 \,,
\end{align}
where the 2-particle state is denoted by $\left| \text{2-ptl st} \right\rangle \,.$
To find more precise interpretations 
will be the subject of future investigation.

\section{Conclusion and Outlook}

In this note, we observed that quench-induced state transition of quantum mechanical system
can be used to capture the excited state spectra of abelian lattice gauge theory in $(1+1)$-dimensions. We introduced two distinct types of gauge-invariant quenches and observed the low-energy excited spectra can be read off
from the loss in the vacuum persistent probability at specific frequency.

The analysis in Section \ref{sec:setup} further allows the estimation of how many controlled-Z (CZ) or CNOT is necessary to identify the excited state at a particular energy level given probability threshold $P_\textrm{th}$ we defined in the paragraph under \eqref{transition-probability}.
The number of the controlled gates at each Trotterized step
depends quadratically on the number of qubits, as shown in \eqref{H_chiral}. Hence,
the total number of controlled gates for 
$M =T /\Delta t_{ST}$ Trotterized steps is bounded below by 
$ \mathcal{N}_{CZ} \sim \mathcal{O} \left( M N^2 \right) > \mathcal{O} \left( \left( \omega \sqrt{ P_\textrm{th} } / \gamma B_p \right)^{\frac{3}{2}}  N^2 \right) \,.$
This depends on the energy scale $\omega$ of the system and the characteristic number of quench $\gamma$.

There are various interesting future directions.
First of all, it should be interesting to implement the simulation on a real quantum device. 
It is of future interest that how the selection rule or symmetry property of the excited states can be confirmed 
by the spectroscopy heuristics described in this note.
Another interesting direction is to compare the computational complexity of our algorithm with those of similar algorithms based on tensor network, which is another powerful approach to the Schwinger model with topological angle \cite{Banuls:2016lkq,Funcke:2019zna,Dempsey:2022nys,Honda:2022edn,Okuda:2022hsq}. 
It would be also illuminating to apply our method to the two-flavor Schwinger model and estimate the mass spectrum of the composite particles whose DMRG simulation was recently done based on different methods \cite{Itou:2023img}.

\section*{Acknowledgement}
\noindent  M. H. is supported by MEXT Q-LEAP, JSPS Grant-in-Aid for Transformative Research Areas (A) JP21H05190 and JSPS KAKENHI Grant Number 22H01222.
M.H. and D.G. are supported by JST PRESTO Grant Number JPMJPR2117.
D.G. is supported by the Basic Science Research Program of the
National Research Foundation of Korea (NRF) under the Ministry of Education in Korea
(NRF-2022R1A6A3A03068148).

\bibliographystyle{utphys}
\bibliography{mybib}


\end{document}